\documentclass[multphys,vecphys]{svmult}

\usepackage{makeidx}         
\usepackage{graphicx}        
\usepackage{multicol}        
\usepackage[bottom]{footmisc}

\makeindex             

\begin{document}

\title*{The stellar populations of E and S0 galaxies as seen with
  \tt{SAURON}}
\author{Harald Kuntschner \inst{1} \and 
Eric Emsellem \inst{2} \and 
Roland Bacon \inst{2} \and 
Martin Bureau \inst{3} \and 
Michele Cappellari \inst{4} \and 
Roger L. Davies \inst{3} \and 
Tim de~Zeeuw \inst{4} \and 
Jes\'us Falc\'on-Barroso \inst{4} \and 
Davor Krajnovi\'c \inst{3} \and 
Richard M. McDermid \inst{4} \and 
Reynier F. Peletier \inst{5} \and 
Marc Sarzi \inst{3}}
\authorrunning{Kuntschner et al.} 
\institute{Space Telescope European Coordinating Facility, Garching,
  Germany \texttt{hkuntsch@eso.org} \and 
Centre de Recherche Astronomique de Lyon, Lyon, France \and
University of Oxford, Oxford, UK \and
Leiden Observatory, Leiden, The Netherlands \and
Kapteyn Astronomical Institute, Groningen, The Netherlands}

%
%
\maketitle

\begin{abstract}
  
  We present selected results from integral-field spectroscopy of 48
  early-type galaxies observed as part of the {\tt SAURON} survey. Maps
  of the H$\beta$, Fe5015, Mg\,$b$\/ and Fe5270 indices in the Lick/IDS
  system were derived for each of the survey galaxies. The metal line
  strength maps show generally negative gradients with increasing
  radius roughly consistent with the morphology of the light profiles.
  Remarkable deviations from this general trend exist, particularly the
  Mg\,$b$\/ isoindex contours appear to be flatter than the isophotes
  of the surface brightness for about 40\% of our galaxies without
  significant dust features. Generally these galaxies exhibit
  significant rotation. We infer from this that the fast-rotating
  component features a higher metallicity and/or an increased Mg/Fe
  ratio as compared to the galaxy as a whole. 
  
  We also use the line strengths maps to compute average values
  integrated over circular apertures of one effective radius, and
  derive luminosity weighted ages and metallicities. The lenticular
  galaxies show a wide range in age and metallicity estimates, while
  elliptical galaxies tend to occupy regions of older stellar
  populations.


\end{abstract}

\section{The {\tt SAURON} survey}
\label{secc:sauron}

We are carrying out a survey of the dynamics and stellar populations of
72 representative nearby early-type galaxies and spiral bulges based on
measurements of the two-dimensional kinematics and line strengths of
stars and gas with {\tt SAURON}, a custom-built panoramic
integral-field spectrograph for the William Herschel Telescope, La
Palma (Bacon et al. \cite{bac01}). The goals and objectives of the {\tt
  SAURON} survey are described in de~Zeeuw et al. \cite{deZ02}, which
also presents the definition of the sample. The full maps of the
stellar kinematics for the 48 elliptical (E) and lenticular (S0)
galaxies are given in Emsellem et al. \cite{em04}. The morphology and
kinematics of the ionised gas emission are presented in Sarzi et al.
\cite{sar05}. Here we summarize selected results of the absorption line
strength measurements, which are described more fully in Kuntschner et
al. \cite{kun06}.

\section{Mg\,$b$\/ isoindex contours versus isophotes}
\label{secc:mgb_cont}
One of the most interesting aspects of integral-field spectroscopy is
the capability to identify two-dimensional structures. For the first
time we can use this in connection with line strength indices and
compare isoindex contours with the isophotal shape. One might expect
the index to follow the light in slowly-rotating giant elliptical
galaxies, since the stars are dynamically well mixed. However, one
could imagine that dynamical substructures with significantly different
stellar populations could leave a signature in line strength maps which
are sensitive to e.g., metallicity.

\begin{figure}
  \centering
  \includegraphics[height=8.3cm]{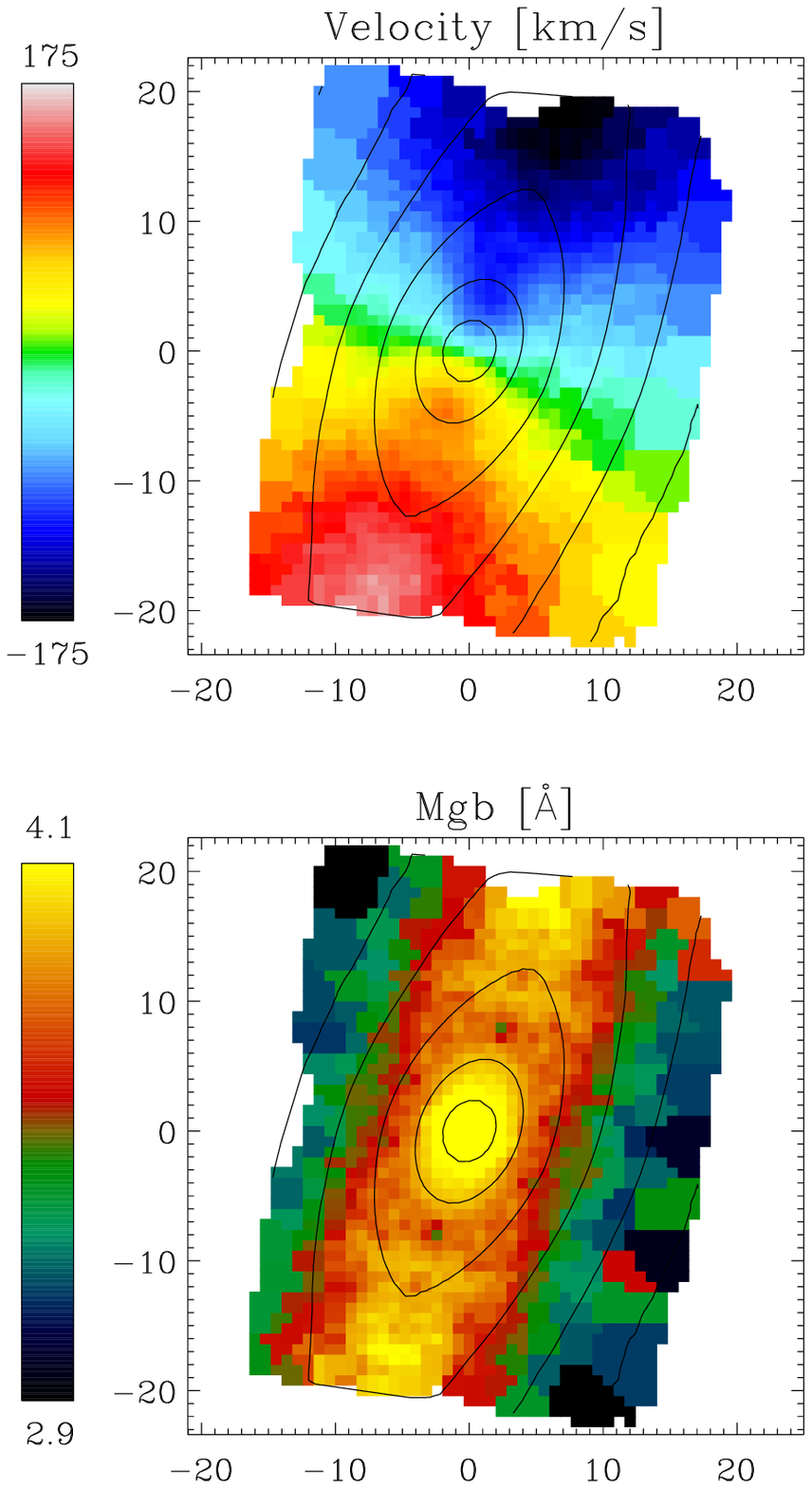}
  \includegraphics[height=8.3cm]{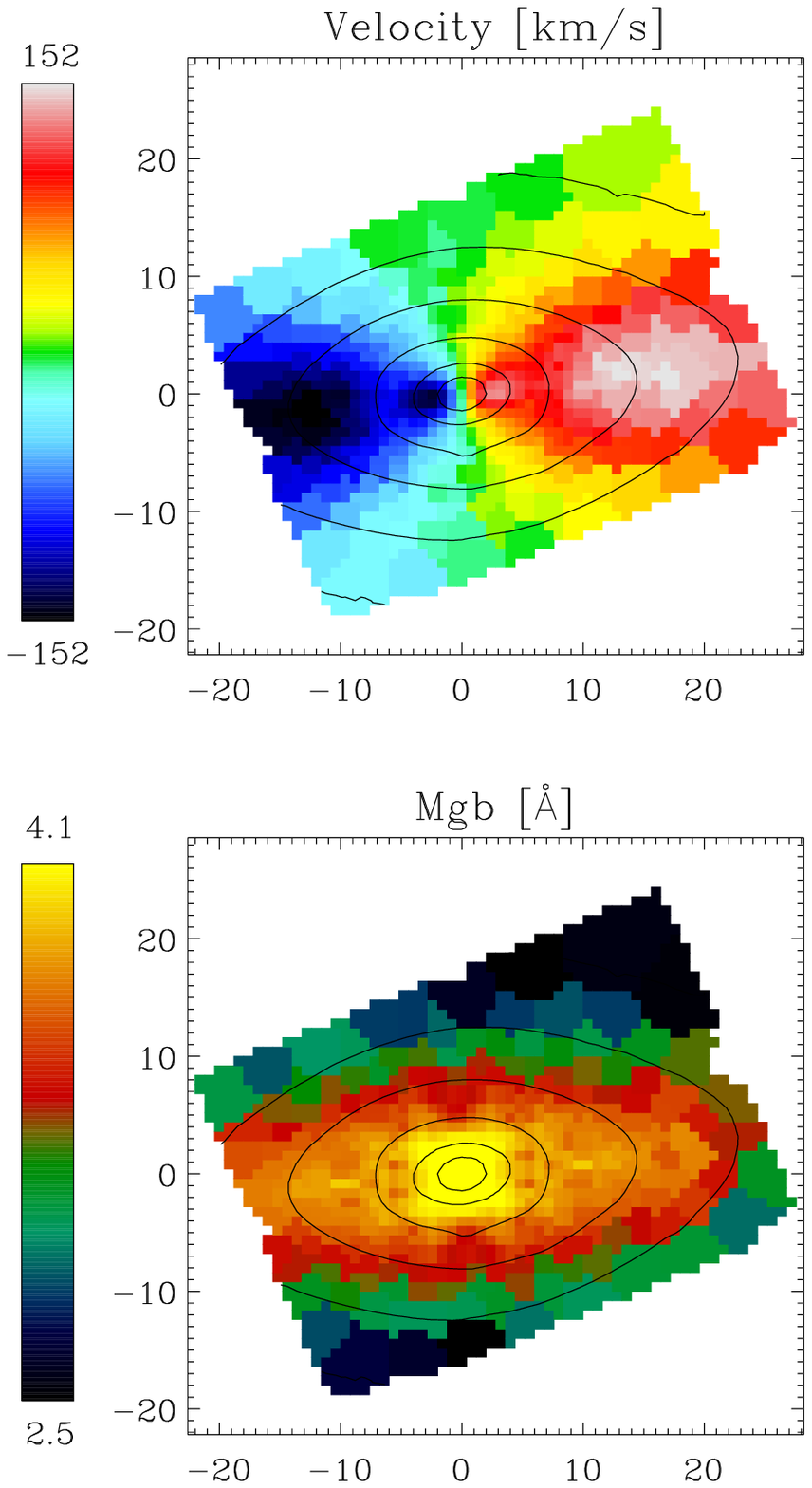}
\caption{The velocity maps and Mg\,$b$\/ maps are shown for NGC\,4570
  (left) and NGC\,4660 (right). Both galaxies show fast-rotating
  components along the direction of enhanced Mg\,$b$\/ strength. For
  presentation purposes the Mg\,$b$\/ maps are symmetrized. The x- and
  y-axis are given in arcsec; North is up and East to the left.}
\label{figg:mgb_n4570}       
\end{figure}

\begin{figure}
  \centering \includegraphics[height=6.0cm]{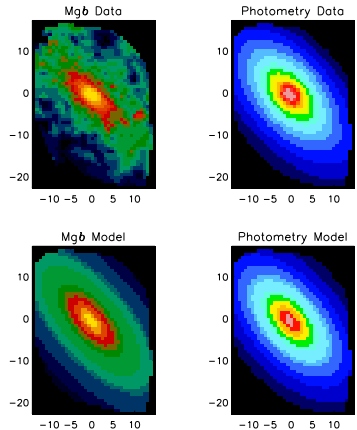}
\caption{{\em Top row:}\/ The interpolated Mg\,$b$\/ map and the
  reconstructed image of NGC\,3377. {\em Bottom row:}\/ The elliptical
  models with constant position angle and ellipticity fitted to the
  Mg\,$b$\/ map and the reconstructed image. The best fitting
  ellipticity for the isophotes and the Mg\,$b$\/ map is
  $0.473\pm0.003$ and $0.573\pm0.022$, respectively. The x- and y-axis
  are given in arcsec; North is up and East to the left.}
\label{figg:mgb_cont_ex}       
\end{figure}

The Mg\,$b$\/ index is the most stringent feature in our survey and
potential differences between isoindex contours and isophotes should be
most apparent. Indeed we find a number of galaxies in our survey where
the isoindex contours clearly do not follow the isophotes, e.g.,
NGC\,4570 and NGC\,4660 (see Fig.~\ref{figg:mgb_n4570}). Both
galaxies show fast-rotating components along the direction of enhanced
Mg\,$b$\/ strength. In order to further study such a possible
connection we determine the best fitting simple, elliptical model for
each of the reconstructed images and Mg\,$b$\/ maps in our survey
(excluding all galaxies with significant dust absorption). For an
example of this procedure see Fig.~\ref{figg:mgb_cont_ex}.

\begin{figure}
  \centering \includegraphics[height=6.5cm]{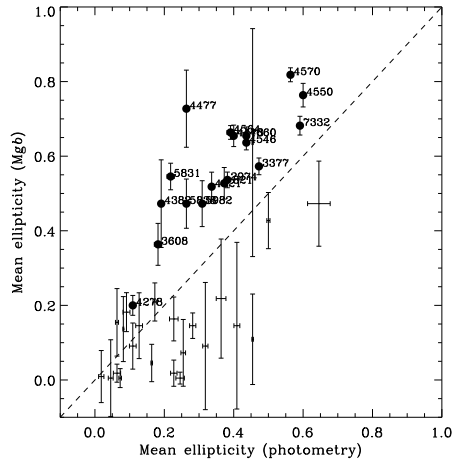}
\caption{Comparison of the average ellipticity of constant Mg\,$b$\/
  strength with the best fitting elliptical model of the isophotes.
  Errors are evaluated by a Monte-Carlo simulation. All galaxies which
  are more than $2\sigma$ above the one-to-one line are indicated by
  filled circles and their NGC numbers.}
\label{figg:mgb_cont}       
\end{figure}

The final results of the ellipse fitting are shown in
Fig.~\ref{figg:mgb_cont}. Applying a $2\sigma$ error cut, 16 out of
41 galaxies appear to have more flattened Mg\,$b$\/ contours than the
isophotes. Most of these galaxies show a high degree of rotational
support in the direction of enhanced Mg\,$b$\/ strength. Thus, the
flattened Mg\,$b$\/ distribution suggests that the fast-rotating
components in these galaxies exhibit a stellar population different
from the main body. The enhanced Mg\,$b$\/ strength can be interpreted
to first order as higher metallicity and/or increased [Mg/Fe] ratio.

\section{Stellar populations}

In order to provide a global measurement of the stellar populations for
each galaxy we derive a central averaged spectrum from all data
available within one effective radius $R_e$. Since our line strength
maps do not always cover the full area of one effective radius we apply
aperture corrections. However, even the galaxies with the smallest
coverage (NGC\,4486 and NGC5846) feature line strength data out to
about 30\% of $R_e$ (corresponds to a field coverage of $\approx
60^{\prime \prime}$) and corrections remain small ($<8\%$). The median
coverage of the line strength maps in our survey is $0.8 R_e$

Estimates of the luminosity weighted age and metallicity of early-type
galaxies can be inferred from index-index diagrams. Generally a
metallicity sensitive index is plotted against an age sensitive index.
In order to minimize the influence of abundance ratio variations on our
metallicity estimates we use a composite index [MgFe50]$^\prime$. This
index is constructed such that a mean metallicity is measured rather
than a metallicity biased by non-solar abundance ratios.

\begin{figure}
  \centering \includegraphics[height=8.5cm]{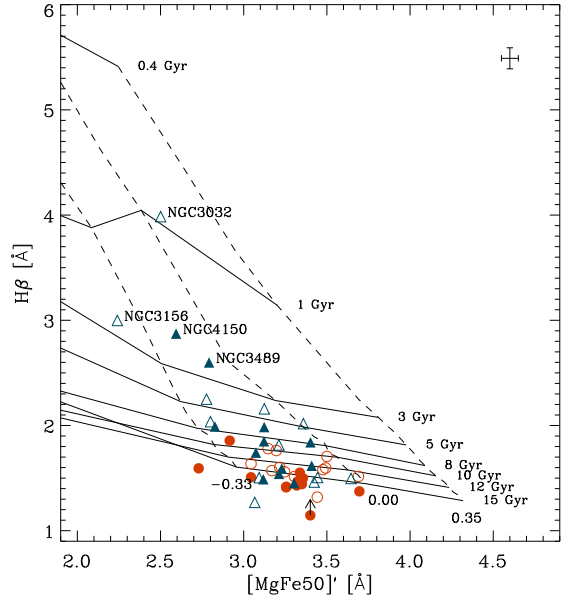}
\caption{Central, one $R_e$ aperture H$\beta$ {\em
    versus}\/ [MgFe50]$^\prime$ line strength measurements. The filled
  and open circles are cluster and field ellipticals, respectively;
  filled and open triangles represent cluster and field S0s,
  respectively. Overplotted are models by Thomas, Maraston \& Bender
  \cite{TMB03}. The error bar in the right upper corner gives the mean
  offset error to the Lick/IDS system. }
\label{figg:mgfe50hb}       
\end{figure}

Fig.~\ref{figg:mgfe50hb} shows a H$\beta$ {\em vs} [MgFe50]$^\prime$
diagram. Overplotted are the model predictions from Thomas, Maraston \&
Bender \cite{TMB03}. The lenticular galaxies in our sample show a wide
range in age and metallicity estimates. The elliptical galaxies
however, tend to occupy regions of older stellar populations.  There is
weak evidence that ellipticals in the field may on average have
experienced some ``frosting'' of young stars yielding lower luminosity
weighted age estimates. It is worth noticing that luminosity weighted
metallicities rarely exceed solar metallicity when taking an average
over one effective radius.

\section{Conclusions}
\label{secc:conclusion}

The results presented here are the outcome of the first comprehensive
survey of the line strength distributions of nearby early-type galaxies
with an integral-field spectrograph. This data set demonstrates that
many nearby early-type galaxies display a significant and varied
structure in their line strength properties. This structure can be of
subtle nature as seen in the deviations of Mg\,$b$ isoindex contours
compared to the isophotes of galaxies.

The 2D coverage of the line strength allows us to connect the stellar
populations with the kinematical structure of the galaxies and thus
further our knowledge of the star-formation and assembly history of
early-type galaxies.

%
%
%



\printindex
\end{document}